\documentclass{PoS}

\PoS{PoS(BDMH2004)}

\usepackage{epsfig}

\title{Extragalactic Background Light: new constraints from the study of the photon-photon absorption on blazar spectra}

\ShortTitle{Extragalactic Background Light and photon-photon absorption}

\author{\speaker{Michela Mapelli}\thanks{We thank A. Celotti, L. Costamante, S. Inoue and T. Matsumoto for useful discussions.}\\ 
        SISSA, Italy\\
        E-mail: \email{mapelli@sissa.it}}

\author{Ruben Salvaterra\\
        SISSA, Italy\\
        E-mail: \email{salvater@sissa.it}}

\author{Andrea Ferrara\\
        SISSA, Italy\\
        E-mail: \email{ferrara@sissa.it}}

\abstract{The study of the Extragalactic Background Light (EBL) is crucial to understand many astrophysical problems (as the formation of first stars, the evolution of galaxies and the role of dust emission). At present, one of the most powerful ways to put constraints on EBL is represented by the study of the photon-photon absorption on gamma-ray spectra of TeV blazars. Adopting this method, we found that, if the only contribution to the optical and Near Infrared (NIR) background is given by galaxies, the spectrum of the blazar H~1426+428 cannot be fitted. To reproduce the observational data of H~1426+428 a Near Infrared excess with respect to galaxy counts is required, with amplitude consistent with both the Matsumoto et al. (2000) data with Kelsall's model of zodiacal light (ZL) subtraction and the DIRBE data with Wright's model of ZL subtraction. The derived constraints on the optical EBL are weaker, because the experimental errors on blazar data are still bigger than the differences among various optical EBL models. In the mid-infrared the {\it SPITZER} measurement of $\nu{}I_{\nu{}}$=2.7 nW m$^{-2}$ sr$^{-1}$ at 24 $\mu{}$m, provides the best fit of the blazar spectrum. 
}

\FullConference{Baryons in Dark Matter Halos\\
		 5-9 October 2004\\
		 Novigrad, Croatia}

\begin{document}

\newcommand{\q}{\begin{equation}}
\newcommand{\qa}{\begin{eqnarray}}
\newcommand{\qs}{\begin{eqnarray*}}
\newcommand{\nq}{\end{equation}}
\newcommand{\nqa}{\end{eqnarray}}
\newcommand{\nqs}{\end{eqnarray*}}
\newcommand{\ud}{\mathrm{d}}

\section{Introduction}
The study of the  Extragalactic Background Light (hereafter EBL) might provide unique information to understand many crucial astrophysical questions,
including, among others, the first cosmic star formation (Salvaterra \&{} Ferrara 2003), the evolution of  galaxies and  the role of  dust emission (Totani \&{} Takeuchi 2002). In this proceeding we want to show as the study of the photon-photon absorption on blazar spectra can provide a powerful method to improve our knowledge of the EBL. In Sec.~2 we summarize the present  measurements and problems about EBL. In Sec.~3 we shortly describe the photon-photon absorption and our procedure. In Sec.~4 we present our findings and in Sec.~5 we report some conclusions.

\section{The Extragalactic Background Light}
\begin{figure}[!h]
\begin{center}
\epsfig{file=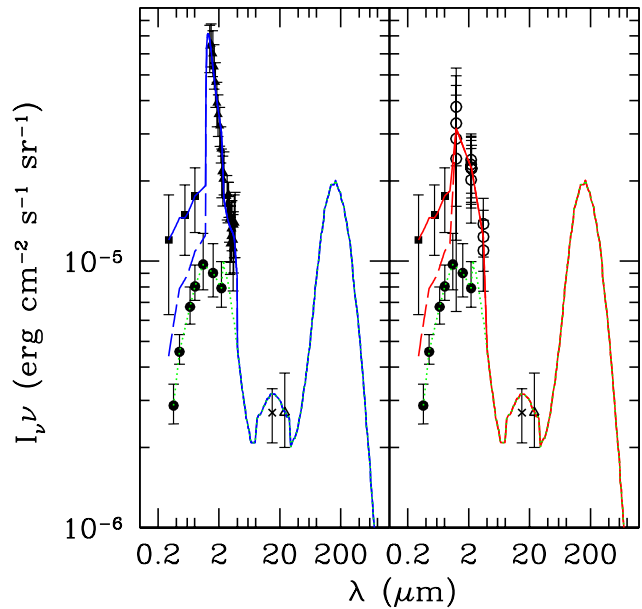, width=.6\textwidth}
\caption{Extragalactic Background Light data and corresponding models. { \bf Left panel:} MK1 ({\it dashed blue line}), MK2 ({\it solid blue line}). The Matsumoto et al. (2000) data are represented with {\it filled triangles}.{\bf Right panel:} DW1 ({\it dashed red line}), DW2 ({\it solid red line}). The Wright (2001) data are represented with {\it open circles}. In both the two panels are shown: C1 ({\it dotted green line}), data from Bernstein et al. ({\it filled squares}), Madau \& Pozzetti ({\it filled circles}), Metcalfe et al. ({\it cross}) and Papovich et al ({\it open triangle}).}
\end{center}
\label{fig1}
\end{figure}
At present, our knowledge of the EBL still presents a lot of uncertainties, both in the optical and in the infrared region. Galaxy counts (Madau \&{} Pozzetti 2000) could entirely represent the optical EBL. But Dube et al. (1979) derived an upper limit of the background flux at $\lambda{}\sim{}5000\,{}\dot{A}$ considerably higher than galaxy counts and, more recently, Bernstein et al. (2002) measured a significant excess at $\lambda{}$=3000, 5500, and 8000 $\dot{A}$ (Fig. 1). On the other hand, this excess can not yet be confirmed, as pointed out by Mattila (2003). An analogue uncertainty about the presence of an excess affects the Near-Infrared region (hereafter NIR). The principal problem of NIR background measurements is represented by the subtraction of Zodiacal Light (hereafter ZL). The Zodiacal Light is the sunlight scattered by interplanetary dust. There are two different models of ZL light subtraction: the Kelsall model (Kelsall et al. 1998), based on the temporal variability of the ZL due to the different amounts of dust encountered by the Earth during its orbit, and the Wright model (Wright 2001 and references here), the maximal one, based on the fact that the Kelsall model seems to be unable to subtract all the ZL  at some wavelengths (for example at 25 $\mu{}$m). In the left panel of Fig. 1 we indicate the Matsumoto data (Matsumoto et al. 2000) of the NIR region, obtained subtracting the ZL contribution through the Kelsall model; while in the right panel we show the COBE/DIRBE data (Wright 2001), where the ZL has been subtracted following the Wright model. Both Matsumoto et al. and Wright data show an excess in the NIR with respect to galaxy counts, even if of different amount.\\
The problem of the Mid and Far-Infrared region is the small number of data. In fact the only recent data are provided by ISOCAM (Metcalfe et al. 2003), measuring a flux of $2.7\pm{}0.62$ nW m$^{-2}$ sr$^{-1}$ at 15$\mu{}$m, and by {\it SPITZER} satellite (Papovich et al. 2004), which gives a background total flux of $2.7^{+1.1}_{
-0.8}$ nW m$^2$ sr$^-1$ at 24 $\mu$m. At present we are waiting for new {\it SPITZER} measurements.
\section{The photon-photon absorption on blazar spectra} 
Because of the present uncertainties about EBL direct measurements, we need an alternative approach to improve our knowledge of the Extragalactic Background Light. This alternative approach can be represented by the study of the photon-photon absorption on blazar spectra, a pair production process involving two photons whose energies, $E$ and $\epsilon{}$, are related by the (approximate) formula $\epsilon{}\sim{}2(m_ec^2)^2/E$ (where $m_e$ is the electron rest mass). This means that, if $E$ is the energy of a photon in the TeV range of a blazar spectrum, the interacting photon should belong to the Infrared Background ($\lambda{}\sim{}2(E/1\textrm{TeV})\,{}\mu{}m)$, where $\lambda{}$ is the wavelength of the photon of energy $\epsilon{}$). This strong relation between the energies of the interacting photons exists because the photon-photon absorption cross section for a given $E$ is a well peaked function of $\lambda{}$ (Fig. 2). 
\begin{figure}[!h]
\begin{center}
\epsfig{file=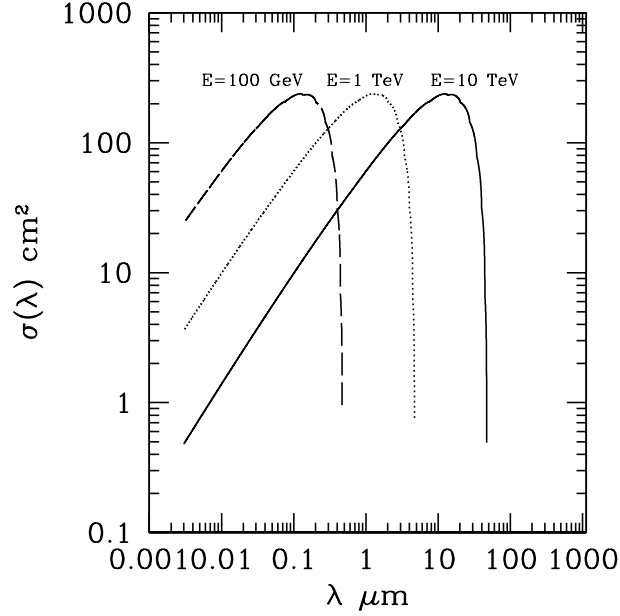, width=.6\textwidth}
\end{center}
\caption{Photon-photon absorption cross section as a function of 
the background photon wavelength, $\lambda$, integrated over the interaction 
angle and over the redshift (with $z_{em}=0.129$, the redshift of H1426+428), 
for three values of the observed energy of the blazar photon: 100 GeV 
({\it dashed line}), 1 TeV ({\it dotted}) and 10 TeV ({\it solid})).}
\label{fig2}
\end{figure}
The full expression of the photon-photon absorption can be written (Madau \& Phinney 1996) as
\q\label{eq:eq1}
\tau{}(E)=\int^{z_{em}}_{0}d{z}\frac{d{l}}{d{z}}\int^{1}_{-1}d{x}\,{}\frac{(1-x)
}{2}\int^{\infty{}}_{\epsilon{}_{th}}d{\epsilon{}}\,{}n(\epsilon{})\,{}\sigma{}(
\epsilon{},E,x),
\nq
where $d{l}/d{z}$ is the proper line element\footnote{We adopt the following 
cosmological parameters:
Hubble constant $H_0$=71 km s$^{-1}$ Mpc$^{-1}$, $\Omega{}_M$=0.27,
$\Omega{}_\Lambda{}$=0.73, which are in agreement with the recent WMAP 
determination (Spergel et al. 2003).}
$\frac{d{l}}{d{z}}=\frac{c}{H_0}\left[(1+z)\,{}{\cal E}(z)\right]^{-1}$,
$c$ is the speed of light and\\
${\cal E}(z)=\left[\Omega{}_M(1+z)^3+\Omega{}_\Lambda{}+(1-\Omega{}_\Lambda{}-
\Omega{}_M)(1+z)^2\right]^{1/2}$.
In eq. \ref{eq:eq1}, $x\equiv{}\cos{\theta{}}$, $\theta{}$ being the angle between the directions of the two interacting photons. $E=E_0\,{}(1+z)$ is the observed energy of the blazar photon and $\epsilon{}=\epsilon{}_0(1+z)$ is the observed energy of the background photon; $z_{em}$ is the redshift of the considered blazar; finally,
$n(\epsilon)$ is the specific number density of background photons.
The energy threshold for the interaction, $\epsilon{}_{th}$, is defined by
$\epsilon{}_{th}=\frac{2\,{}m_e^2\,{}c^4}{E\,{}(1-x)}$,
where $m_e$ is the electron mass.\\
The photon-photon absorption cross section is given by
\q\label{eq:eq3}
\sigma{}(\epsilon{},E,x)=\frac{3\,{}\sigma{}_T}{16}(1-\beta{}^2)\left[2\,{}\beta{}(\beta{}^2-2)+(3-\beta{}^4)\ln{\left(\frac{1+\beta{}}{1-\beta{}}\right)}\right]
\nq
where $\sigma_T$ is the Thompson cross section and
$\beta{}\equiv{}\left[1-\frac{2\,{}m_e^2\,{}c^4}{E\,{}\epsilon{}\,{}(1-x)\,{}}
\right]^{1/2}$
We integrated numerically the expression \ref{eq:eq1}
 for different EBL models, using a three-dimensional integral based on the method of Gaussian quadratures.

The absorbed spectrum of the blazar ($\left(\frac{d{N}}{d{E}}\right)_{abs}$) can be derived by the unabsorbed one ($\left(\frac{d{N}}{d{E}}\right)_{unabs}$), given the relation:
\q\label{eq:eq6}
\left(\frac{d{N}}{d{E}}\right)_{abs}=e^{-\tau{}(E)}\left(\frac{d{N}}{d{E}}
\right)_{unabs},
\nq
Assuming that the unabsorbed spectrum has a power law shape\footnote{ The assumption that the unabsorbed spectrum follows a power law is valid especially in the case of the blazar we are going to analyze (H 1426+428). For other blazars the unabsorbed spectrum is better fitted by a power law with an exponential cut-off (Konopelko et al. 2003).} ($(d{N}/d{E})_{unabs}\propto{}E^{-\alpha{}}$) and calculating $\tau{}(E)$ for various EBL models, we can derive the absorbed spectrum  and compare it with observational data. Through this procedure we are able to distinguish among various EBL models. We apply this method to 5 different EBL models, reported in Table 1.

\begin{table}
\begin{center}
\begin{tabular}{llll}
\hline
\hline
{\bf Model} & Optical Background & NIRB & MIRB-FIRB\\
\hline

C1          & \multicolumn{2}{c}{Madau \& Pozzetti (2000)}  &  Totani \& Takeuchi (2002) + {\it SPITZER}$^{b}$\\
MK1          & Madau \& Pozzetti (2000)  & Matsumoto et al. (2000) (K)$^{a}$ & Totani \& Takeuchi (2002) + {\it SPITZER} \\
MK2          & Bernstein et al. (2002)  & Matsumoto et al. (2000) (K) & Totani \& Takeuchi (2002) + {\it SPITZER} \\
DW1          & Madau \& Pozzetti (2000) & Wright (2001) (W)$^{a}$ & Totani \& Takeuchi (2002) + {\it SPITZER} \\
DW2          & Bernstein et al. (2002) & Wright (2001) (W) & Totani \& Takeuchi (2002) + {\it SPITZER} \\
\hline
\end{tabular}
\caption{Summary of the considered EBL models.
}
\end{center}
{\footnotesize $^{a}$(K) and (W) indicate the ZL subtraction obtained using Kelsall's model and Wright's model, respectively.}\\
{\footnotesize $^{b}$Totani \& Takeuchi model rescaled to the {\it SPITZER} data (Papovich et al. 2004).}\\
\label{tab_1}
\end{table}
\section{Results}
The first result we obtained adopting our method is that an EBL excess (in the optical or NIR) is required to fit the TeV spectrum of the blazar H~1426+428 (redshift $z=0.129$). In fact adopting the model C1, i.e. assuming that both the optical and the NIR background are represented by galaxy counts, we cannot fit the spectrum of H 1426+428, as you can see from Fig. 3 and from the $\chi{}^2$ test in Table 2.
\begin{figure}[!h]
\begin{center}
\epsfig{file=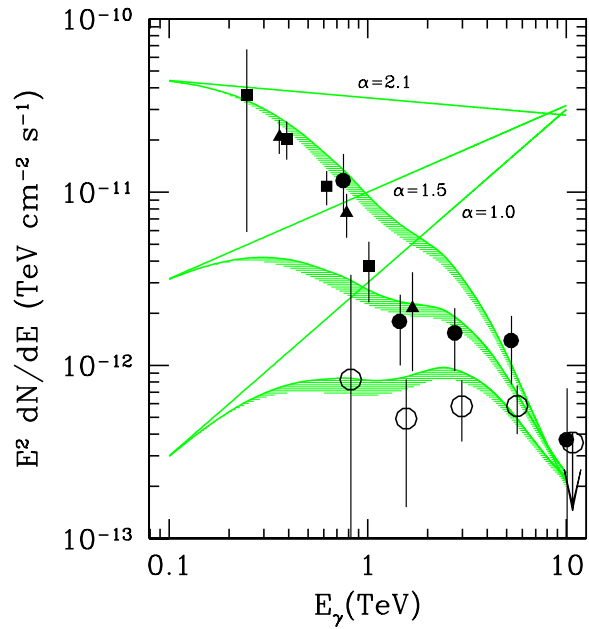, width=.6\textwidth}
\end{center}
\caption{Three different fits to the spectrum of H~1426+428. The
 photon-photon absorption has been calculated assuming the background model C1 (see Table~1).
The shaded area indicates the region between the absorbed spectra obtained with
the upper limit (+$1\sigma{}$) of galaxy counts and that obtained with the lower limit (-$1\sigma{}$) of galaxy counts. The {\it solid line} represents the case C1 (see Table~1).
 The observational data reported here are from CAT  1998-2000 ({\it filled squares}), Whipple 2001 ({\it filled triangles}), HEGRA 1999-2000 ({\it filled circles}), HEGRA 2002 ({\it open circles}). All error bars are at 1 $\sigma{}$ (Aharonian et al. 2003). }
\label{fig3}
\end{figure}


\begin{table}
\begin{center}
\begin{tabular}{llllll}
\hline
\hline
             & C1 & MK1   & MK2   & DW1  & DW2    \\
\hline
 $\chi{}^2$  & 48.00 & 18.69 & 19.71 & 6.98 & 7.41 \\
\hline
\end{tabular}
\end{center}
\caption{$\chi^2$ values  of the considered models.
}
{\footnotesize We calculated the $\chi{}^2$ for each one of the considered models. We based our statistical analysis on 12 observational data of the spectrum of H1426+428 (CAT  1998-2000, Whipple 2001, HEGRA 1999-2000).}
\label{tab_2}
\end{table}


Then we need of an EBL excess. If we assume that this excess is in the NIR region, we can fit the spectrum of H~1426+428 (Fig. 4), both if we calculate $\tau{}(E)$ adopting the Kelsall model of ZL subtraction for the NIR (MK1 and MK2 models) and if we calculate $\tau{}(E)$ adopting the Wright model (DW1 and DW2). The DW1 model gives the best result, considering the $\chi{}^2$ (Table 2); but we cannot exclude also the other 3 cases (DW2, MK1 and MK2). \\
Unfortunately, our results are not conclusive concerning the existence of an optical excess. The reason is, as shown in Fig. 4, that the difference between the results obtained using only galaxy counts for the optical background and those obtained adopting the Bernstein et al. measurements is much smaller than the error bars in the observational data of the blazar spectrum.
\begin{figure}[!h]
\begin{center}
\epsfig{file=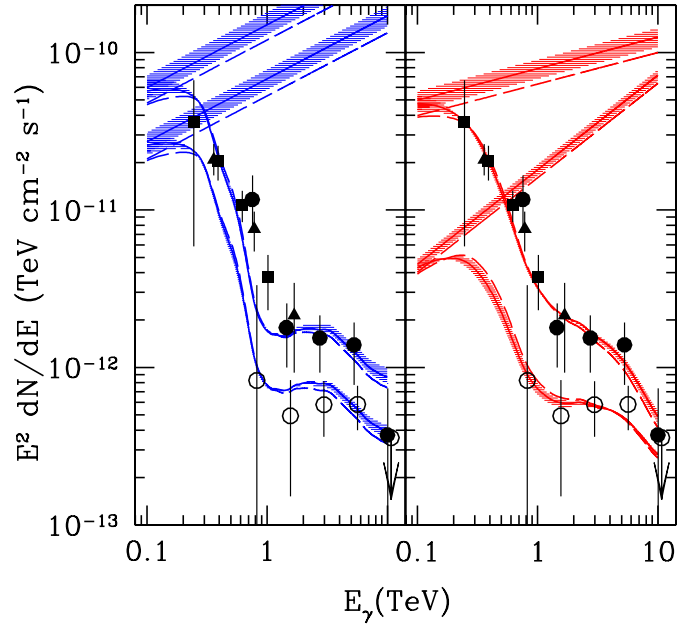, width=.6\textwidth}
\end{center}
\caption{Fit to the H~1426+428 spectrum. The EBL models assumed are: {\bf Left panel:} MK1 ({\it dashed blue line}) and MK2 ({\it solid}); {\bf Right panel:} DW1 ({\it dashed red line}) and DW2 ({\it solid}). Shaded areas refer to $\pm 1\sigma$ errors. the observational data are the same as in Fig.~3.}
\label{fig4}
\end{figure}

Using our method we can also derive the most likely spectral index of the blazar spectrum. In particular, if we use the DW1 model, we obtain for H 1426+428 a spectral index $\alpha{}=1.8$, in good agreement with previous results (Aharonian et al. 2003).

Our method can provide also some constraints on the {\it SPITZER} data, at least under the assumption that the Totani \& Takeuchi (2002) model of Mid and Far Infrared is correct. Let's consider the {\it SPITZER} measurement at 24 $\mu{}$m (total background flux $F=2.7^{+1.1}_{-0.7}\,{}\textrm{nW m}^{-2}\textrm{sr}^{-1}$). We obtain a good fit of the blazar spectrum only adopting the {\it SPITZER} best value; while using the upper or the lower limit of this measurement our results are not in agreement with some observational data~(Fig.~5).
\begin{figure}[!h]
\begin{center}
\epsfig{file=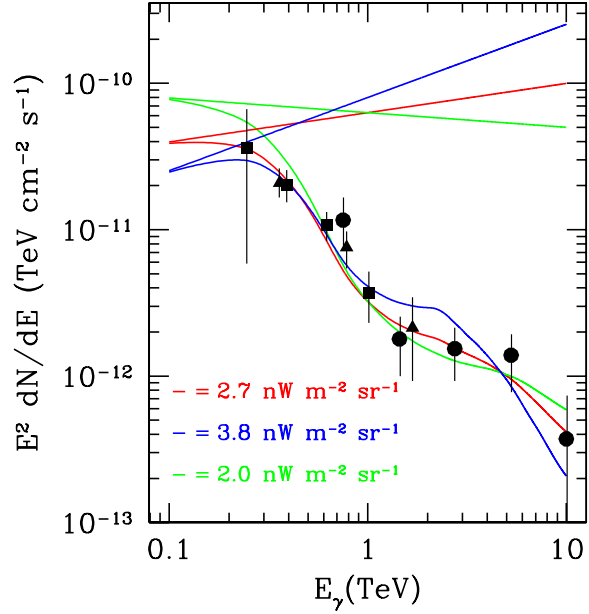, width=.6\textwidth}
\end{center}
\caption{ Fit of the H~1426+428 spectrum. We have assumed an EBL model given by: (a) optical background (0.3-1.2 $\mu{}$m): Madau \& Pozzetti 2000; (b) near infrared background (1.2-4. $\mu{}$m): DIRBE (Wright 2001) with the ZL model of Wright \& Reese (2000); (c) middle and far infrared background : model of Totani \&{} Takeuchi (2002). The model of Totani \& Takeuchi for the Mid and the Far Infrared has been rescaled in the 8-30 $\mu{}$m range assuming different values of the EBL at 24 $\mu{}$m, in particular assuming: 2.7 ({\it red line}),
3.8 ({\it blue line}) and 2 ({\it green line}) nW m$^{-2}$ sr$^{-1}$.  The observational data reported here are from CAT  1998-2000 ({\it filled squares}), Whipple 2001 ({\it filled triangles}), HEGRA 1999-2000 ({\it filled circles}).}
\label{fig5}
\end{figure}
 
\section{Summary}
In conclusion, the most important finding of our work is that a Near Infrared excess with respect to galaxy counts must be present, if we want to explain the photon-photon absorption deduced from blazar spectra. We cannot definitely distinguish between the NIR excess derived adopting the Kelsall model of ZL subtraction and that obtained from the Wright model; but the $\chi{}^2$ analysis favors the latter. This result is crucial for models describing the connection between EBL and Population III stars (Salvaterra \& Ferrara 2003). In fact these models, which explain the EBL in the Near Infrared region as the radiation emitted by high $z$ metal free stars and redshifted in the infrared region, are in agreement with the existence of a NIR excess with respect to galaxy counts.  

Our work presents also a lot of challenges. First of all we need new measurements in the 0.1-1 TeV range of the spectrum of H 1426+428, because this region is crucial to distinguish among various optical background models. These new measurements could be provided by new Cherenkov telescopes, which are already at work (MAGIC, VERITAS, HESS) or will be operating in a few years (GLAST). We need also new spectra of other high redshift blazars, because now our analysis is based on only one blazar (H 1426+428). In particular, there are at least two TeV blazars with redshift similar (PKS 2155-304 at $z=0.117$) or higher (1ES 2344+514 at $z=0.44$) than H1426+428; but the data which are available for these blazars (Roberts et al. 1998, Catanese et al. 1998) are not sufficiently accurate for our analysis. Finally, we need also new {\it SPITZER} data, to improve our knowledge of the Mid and Far Infrared background.

\end{document}